\documentstyle[aps,epsfig,multicol]{revtex}
\def\breakon{\end{multicols}\widetext\vspace{-.2cm}
\noindent\rule{.48\linewidth}{.3mm}\rule{.3mm}{.3cm}\vspace{.0cm}}

\renewcommand{\vec}[1]{{\mathbf #1}}

\def\breakoff{\vspace{-.2cm}
\noindent
\rule{.52\linewidth}{.0mm}\rule[-.27cm]{.3mm}{.3cm}\rule{.48\linewidth}{.3mm}
\vspace{-.3cm}
\begin{multicols}{2}
\narrowtext}

\newcommand{\be}{\begin{equation}}
\newcommand{\ee}{\end{equation}}
\newcommand{\bea}{\begin{eqnarray}}
\newcommand{\eea}{\end{eqnarray}}
\newcommand{\HH}{{\cal H}}

\newcommand{\la}{\langle}
\newcommand{\ra}{\rangle}

\newcommand{\lp}{\left(}
\newcommand{\rp}{\right)}
\renewcommand{\phi}{\varphi}
\renewcommand{\vec}[1]{{\bf #1}}

\begin{document}
\title{Atom-molecule coexistence and collective dynamics near a Feshbach resonance of cold fermions}
\author{R.\,A. Barankov and  L.\,S. Levitov}
\address{Department of Physics,
Massachusetts Institute of Technology, 77 Massachusetts Ave,
Cambridge, MA 02139}

\maketitle
\begin{abstract}
Degenerate Fermi gas interacting with molecules near Feshbach resonance
is unstable with respect to formation of a mixed state 
in which atoms and molecules coexist as a coherent superposition.
Theory of this state is developed using a mapping to the Dicke model,
treating molecular field in the single mode approximation. 
The results are accurate in the strong coupling
regime relevant for current experimental efforts.
The exact solution of the Dicke model is exploited to study
stability, phase diagram, and
nonadiabatic dynamics of molecular field in the mixed state.
\end{abstract}


\begin{multicols}{2}

\narrowtext
Feshbach resonance scattering~\cite{Inouye98,Loftus02,Jochim02,Marte02},
at which pairs of atoms can bind to form molecules at the same energy,
has been used to demonstrate new coherence phenomena in cold atom systems.
Those include, notably,
the reversible  coherent atom-molecule
transitions~\cite{Donley02,Kokkelmans02} which can be
accompanied by the Bose-Einstein condensation 
of molecules~\cite{Claussen02,Joshim030,Durr04}.
Recently, in search of fermionic condensation, the focus shifted to
Feshbach resonance in cold fermion systems~\cite{Regal04,Kinast04,Zwierlein04}.

The physics near the resonance
in a macroscopic system
is sensitive to the effects of quantum statistics.
In particular, at positive detuning from 
the resonance molecules can coexist with
fermions~\cite{Timmermans01,Holland01,Falco04,Bruun04}, 
stabilized by Pauli blocking
of molecular decay 
into the states below the Fermi level.

The stability and properties of the mixed state depend
on the interaction effects.
Below we argue that the interactions greatly
enhance stability of the atom-molecule
mixture, and lead to molecules and atom pairs hybridizing
to form a coherent state.
We address the problem of molecules interacting
with atoms by mapping it onto the Dicke problem~\cite{Dicke54} of two-level
systems coupled to a Bose field. 
This problem, being exactly solvable~\cite{Hepp73}, allows to describe
the experimentally relevant regime of strong coupling.
In the Feshbach resonance case, 
the two-level systems represent fermion pair states 
which can be occupied or empty, while
the Bose field represents molecules.

The coupling to molecules at 
positive detuning from Feshbach resonance enhances pairing
interaction between fermions, which is expected to stimulate BCS 
superfluidity~\cite{Bohn00,Timmermans01,Holland01,Falco04,Bruun04,Ohashi02,Milstein02}.
In addition,
as noted by Timmermans {\it et al.}~\cite{Timmermans01} and 
Holland {\it et al.}~\cite{Holland01}, 
the strong coupling BCS condensation,
with the critical temperature up to a fraction of $E_F$,
may depend on the presence of molecular field.
This conclusion was strengthened by
a microscopic analysis carried out by Ohashi and Griffin~\cite{Ohashi02},
by Milstein {\it et al.}~\cite{Milstein02}, 
who refine the approach of Ref.~\cite{Holland01}.
Bruun and Pethick~\cite{Bruun04} studied
noncondensed molecules coexisting with
the Fermi gas at positive detuning 
using an effective theory of strong coupling formulated in terms
of low energy parameters.
It was noted that strong many-body effects exist
even for detuning well above the Fermi energy.
The important role of molecular field at positive detuning
has been reemphasized recently
by Falco and Stoof~\cite{Falco04}, who 
argued that a BCS-BEC crossover 
takes place in this region.

In this article, we focus on the effects of molecule-atom hybridization
and develop an approach allowing to handle this problem
in the strong coupling regime. This is of interest, since the experiment
deals with systems where the atom-molecule coupling,
measured in the Fermi energy
units, is very large.
We will see that 
molecule-atom mixing occurs in this situation
in the range of detuning much larger than the Fermi energy, i.e. on the energy
scale very different from that of fermionic condensate. 
The energy scale for the latter,
set by the pairing interaction strength, 
expected to reach $0.2\,E_F$ at best\cite{Milstein02}, 
is much smaller than the atom-molecule interaction. 
Thus accurate results can be obtained with the help of a simple analysis
which ignores direct pairing interaction between fermions,
and relies on the exact solution of the atom-molecule 
dynamics.

Below we analyze stability of fermions 
with respect to molecule formation,
and obtain a phase digram.
There is a fairly wide region
around the resonance, spanning both positive and negative detuning, 
were atoms and molecules
coexist, forming a coherent state. At strong coupling,
this region has width of the order of $g^2 n/E_F$,
a quantity which different estimates~\cite{Falco04,Bruun04}
put between few tens and few hundred $E_F$
for current experiments~\cite{Regal04,Kinast04,Zwierlein04}. 
Also, we exploit the Dicke problem solution
to obtain nonlinear oscillations
of the molecular field,
in which population coherently oscillates between
molecular and atomic components.
The results of stability analysis are 
verified by comparing to 
the exact solution 
and to the thermodynamic ground state properties.

We consider the problem of
a Fermi gas interacting with molecules in a single mode approximation
which takes into account only the lowest energy molecular state:
\be\label{H_feshbach}
\HH=\sum_{p,\alpha} \frac{p^2}{2m} a^+_{p,\alpha}a_{p,\alpha}
+g\sum_p\lp b^+c_p + {\rm h.c.}\rp
+\omega b^+b
\ee
with $a_p$, $a_p^+$ and $b$, $b^+$ the atom and molecule operators,
$\alpha$ the fermion spin, and $\omega$ the energy of a molecule.
The atom pair creation and annihilation operators
$c_p=\frac1{\sqrt{2}}\lp a_{-p\downarrow}a_{p\uparrow}
+a_{p\downarrow}a_{-p\uparrow}\rp$,
$c_p^+=\frac1{\sqrt{2}}\lp a_{p\uparrow}^+a_{-p\downarrow}^+
+a_{-p\uparrow}^+a_{p\downarrow}^+\rp$
describe
pairs of fermions in a spin singlet state
that undergo conversion into molecules at Feshbach resonance.
The approximation (\ref{H_feshbach})
is justified by the analysis below which finds
that the energy gained by 
a formation of a mixed atom-molecule state,
with all
molecules occupying one state, is large compared to $E_F$,
which allows to limit consideration to
a single molecular state.

The utility of the single mode approximation (\ref{H_feshbach})
is that it turns a difficult many-body problem
into a well-known exactly solvable
problem. The mapping is achieved by identifying
the pair operators $c_p$, $c_p^+$ with
pseudospin Pauli operators~\cite{Anderson58}
\be
c_p=\sigma_p^-\equiv \frac12(\sigma_p^x-i\sigma_p^y)
,\quad
c_p^+=\sigma_p^+\equiv \frac12(\sigma_p^x+i\sigma_p^y)
\ee
and noting that their product gives the particle number operator
$n_p=a_p^+a_p$
in the subspace of the many-body Hilbert space in which both
states $p$ and $-p$ are simultaneously filled or empty,
$2 c_p^+c_p\equiv n_p+n_{-p}=0,2$. More formally, defining
$\sigma_p^z=[\sigma_p^+,\sigma_p^-]$, one verifies that
the standard Pauli spin commutation relations hold:
\be
[\sigma_p^+,\sigma_p^z]=-2\sigma_p^+
,\quad
[\sigma_p^-,\sigma_p^z]=2\sigma_p^-
\ee
This enables one to bring the Hamiltonian (\ref{H_feshbach})
to the form containing the
spin variables only,
\be\label{H_dicke}
\HH={\sum_p}'\lp \frac{p^2}{2m}\sigma_p^z + g b\sigma_p^+ + gb^+\sigma_p^-\rp
+\omega b^+b
\ee
where the sum is taken over singlet pair states with momenta $p$ and $-p$.
We note that the states with $n_p+n_{-p}=1$, with only one of the
$p$ and $-p$ particle states filled and the other one empty,
are decoupled and do not participate in the dynamics
defined by (\ref{H_dicke}).
The reason for this decoupling is that these states
have not enough particles to form a molecule, but
also one particle too many to
contribute to molecule dissociation.

The spin-boson problem (\ref{H_dicke}) is the
Dicke model of quantum optics~\cite{Dicke54,Hepp73,sculzub}.
Hepp and Lieb\cite{Hepp73} found that
the Hamiltonian (\ref{H_dicke}) is integrable,
and constructed exact many-body states. 
Besides the total particle number
\be
N=2b^+b + \sum_p \lp 1+\sigma_p^z \rp
\ee
there are also infinitely many nontrivial conserved quantities
underpinning the exact solubility.

The problem (\ref{H_dicke})
resembles in many ways the BCS pairing problem.
The latter is also integrable, which allows to obtain
the full energy spectrum, and to construct
nontrivial conserved quantities in a closed form~\cite{Cambiaggio97}. 
In fact, the above pseudospin trick
has its roots in the BCS problem, where it was invented by
Anderson~\cite{Anderson58} 
as an interpretation of Bogoliubov mean field theory.

Here we employ the Hamiltonian (\ref{H_dicke})
to assess stability of the Fermi gas
with respect to molecule formation. The spin dynamics described by
(\ref{H_dicke}) is of the Bloch form,
$\dot\sigma=i[\HH,\sigma]=2\vec h_p\times\sigma$,
with an effective magnetic field $\vec h_p=(gb',gb'',p^2/2m)$, where
$b=b'+ib''$ is a c-number describing the molecular state
in macroscopic limit.

The Bloch equations of motion for
the spin components $\sigma_p^\pm$, $\sigma_p^z$ and $b$
take the form
\bea\label{eq:bloch1_oper}
&& i\dot\sigma_p^+ = -(p^2/m)\sigma_p^+ + gb\sigma_p^z
,\quad
i\dot\sigma_p^- = (p^2/m)\sigma_p^- - gb\sigma_p^z
\\\label{eq:bloch2_oper}
&& i\dot\sigma_p^z = 2gb \sigma_p^+ - 2gb^\ast \sigma_p^-
,\quad
i\dot b = g{\sum_p}'\sigma_p^- + \omega b
\eea
From a mathematical standpoint, Eqs.(\ref{eq:bloch1_oper}),(\ref{eq:bloch2_oper})
describe collective dynamics of a Bloch spin $1/2$ ensemble,
with the coupling between the spins provided by the
`magnetic field' $\vec h_p$
transverse components which depend on the spin variables
via an equation for $b$. Physically, the transverse spin components
$\sigma_p^\pm$ characterize coherence between
the filled and unfilled pair state, while $\sigma_p^z$
describes the number of pairs.

Since the field $b$ is a c-number, the operator equations
(\ref{eq:bloch1_oper}),(\ref{eq:bloch2_oper})
are linear, and thus 
the spin operators expectation values 
are subject to evolution equations
of the form identical to (\ref{eq:bloch1_oper}),(\ref{eq:bloch2_oper}).
In the absence of molecules, we have $b=0$, and all the spins are aligned
in the $\pm z$ direction, with probabilities determined by occupation
of pair states:
$\la \sigma_p^z\ra =p_\uparrow-p_\downarrow=n_p^2-(1-n_p)^2=2n_p-1$,
where $n_p=(e^{\beta(p^2/2m-\mu)}+1)^{-1}$ in thermal equilibrium.
This state, containing only fermions but no molecules,
$\la b\ra=\la \sigma_p^\pm\ra=0$, is stationary for the problem
(\ref{eq:bloch1_oper}),(\ref{eq:bloch2_oper}). 

To assess stability with respect to molecule formation,
we linearize Eqs.(\ref{eq:bloch1_oper}),(\ref{eq:bloch2_oper}),
introducing
$\delta\sigma_p^-,\delta b\propto e^{-i\lambda t}$,
$\delta\sigma_p^+,\delta b^\ast\propto e^{i\lambda^\ast t}$.
From the coupled linear equations for $\delta\sigma_p^-$
and $\delta b$ we obtain the eigenvalue equation
\be\label{eq:lambda_bare}
\lambda = \omega + g^2\sum_p\frac{\la \sigma_p^z\ra}{p^2/m-\lambda}
\ee
To make the formally divergent sum over $p$ well-behaved, following
Ohashi and Griffin~\cite{Ohashi03}, 
we renormalize $\omega$
by subtracting the term $\delta\omega=g^2\sum_p(p^2/m)^{-1}$.
The shift $\omega \to \omega-\delta\omega$ brings the position
of the Feshbach resonance to $\omega=0$ for zero particle density,
while Eq.(\ref{eq:lambda_bare}) transforms to
\be\label{eq:lambda}
\lambda = \omega + g^2\sum_p\lp \frac{2n_p-1}{p^2/m-\lambda}
+\frac1{p^2/m}\rp
\ee
with the sum now converging at large $p$.

The solution of Eq.(\ref{eq:lambda}) can be real or complex,
depending on the value of $\omega$.
Complex-valued $\lambda=\lambda'+i\lambda''$
indicates an instability, with $\lambda''$ describing the
instability growth rate. Numerical analysis
of the solutions of Eq.(\ref{eq:lambda}) 
and simple analytic arguments 
reveal that
(i) the real part $\lambda'$ is a monotonic function
of $\omega$;
(ii) the instability occurs in an interval
$\omega_0<\omega<\omega_1$ with the threshold values
$\omega_{0,1}$ being a function of temperature.

\begin{figure}[t]
\centerline{
\begin{minipage}[t]{3.5in}
\vspace{0pt}
\centering
\includegraphics[width=3.5in]{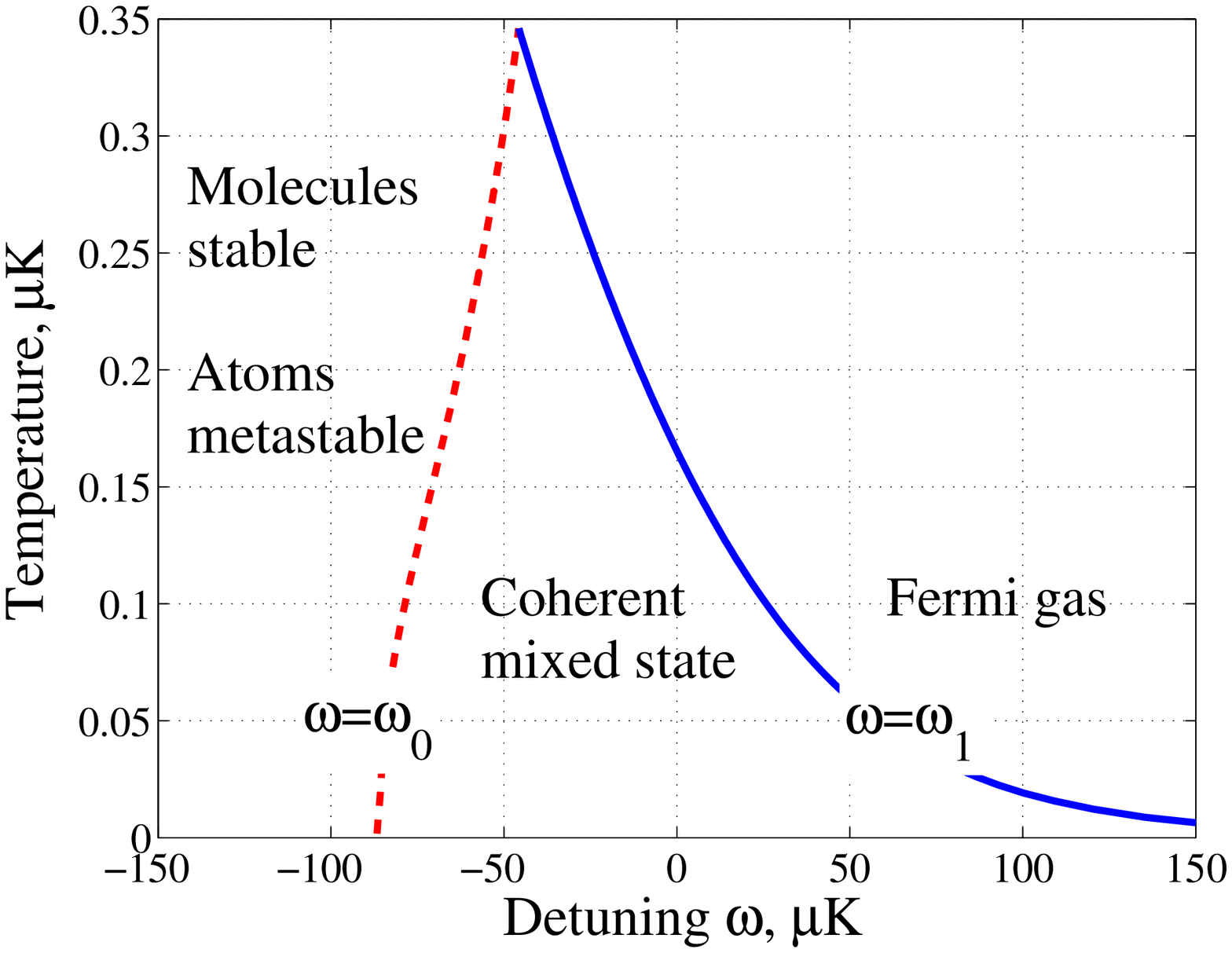}
\end{minipage}
\hspace{-1.65in}
\begin{minipage}[t]{1.6in}
\vspace{0.0in}
\centering 
\includegraphics[width=1.5in]{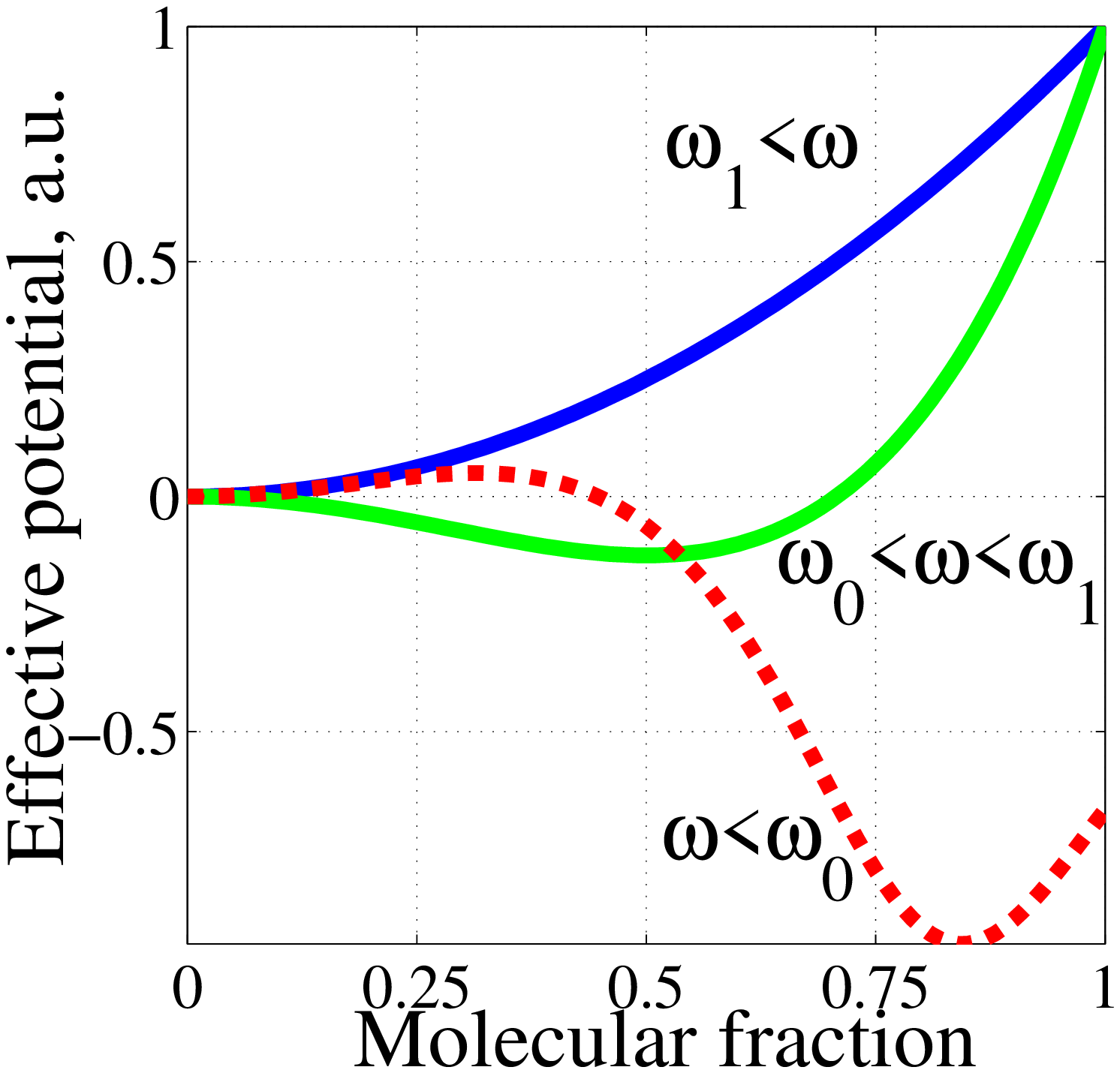}
\end{minipage}
} \vspace{0.15cm} 
\caption[]{Phase diagram of coupled atom-molecule system 
obtained from Eq.(\ref{eq:boundary}) for 
$^{40}K$ system~\cite{Regal04} 
at particle density $n\approx 1.8\times 10^{13}{\rm cm^{-3}}$, 
Fermi energy $E_F=0.35\mu{\rm K}$,
and coupling strength $g^2 n/E_F\approx 60\mu{\rm K}$.
(The coupling was estimated using the microscopic theory of Feshbach
resonance developed by Falco and Stoof~\cite{Falco04},
applied to the conditions of the JIlA experiment~\cite{Regal04}).
\emph{Inset:} Effective potential schematic illustrating the behavior
in the three regions.
} \label{fig:Phase-diagram}
\end{figure}

The values $\omega_{0,1}$ can be inferred by noting that the complex
$\lambda$
becomes real at $\omega=\omega_{0,1}$, which 
gives the condition $\lambda''=0$.
When does Eq.(\ref{eq:lambda}) admit real solutions?
This is possible 
for $\lambda\le 0$, as well as
for $\lambda=2\mu$, 
since $2n_p-1$ changes sign at $p=p_F$.
(For all positive $\lambda$ except $\lambda=2\mu$ the residue
$\la \sigma_p^z\ra = 2n_p-1$ generates a finite imaginary part of $\lambda$.)
With $\lambda=0,\,2\mu$ one obtains
\bea\label{eq:boundary}
&&\omega_0= - g^2\sum_p\frac{2n_p}{p^2/m}
\\
&&\nonumber
\omega_1=2\mu+ g^2\sum_p\lp \frac{1-2n_p}{p^2/m-2\mu}-\frac1{p^2/m}\rp
\eea
This indicates that atoms are stable at $\omega>\omega_1$, metastable
at $\omega<\omega_0$, and 
at $\omega_0<\omega<\omega_1$ can exist only in a state
coherently mixed with the molecules (Fig.~\ref{fig:Phase-diagram}). 
We note that, since $\omega_0<0$ and $\omega_1>2\mu$,
coexistence is favored by
interaction. Moreover, 
at strong interaction, 
the detuning range where coexistence takes place
becomes very large:
$\Delta\omega \simeq g^2 n / E_F \gg E_F$.

The upper temperature at which 
$\omega_0=\omega_1$ is determined by the condition
$\mu(T)=0$. 
For a two-species Fermi gas of total 
particle density $n$ one has
$n=2\sum_p n_p(\mu=0)=0.0972
(m/\beta)^{3/2} $
which gives
$T_\ast =  
0.9885\,E_F$.
Interestingly, at low temperature $T\ll T_\ast$, 
the instability is pushed to higher detuning,
$\omega_1=2\mu+g^2\nu \ln(\mu/T)$,
due to a BCS-like log
divergence at the Fermi level $p=p_F$.

It is instructive to look at 
the JILA experiment parameters (Fig.~\ref{fig:Phase-diagram}). 
The estimate of coupling 
$\Delta \omega = g^2 n/E_F \simeq 60\mu{\rm K} \approx 8\,{\rm MHz}$
gives a typical 
energy gained by the system 
via molecules and atom pairs hybridization,
which is much larger than $E_F$. This leads
to pair size in the mixed state
$\sim \hbar/(2m\Delta\omega)^{1/2}$
much smaller than fermion wavelength $p_F^{-1}$.
This indicates that the kinetic energy of atoms \emph{and molecules}
does not play a significant role, justifying
the single mode approximation.

Nonlinear dynamics at instability can be found
with the help of the
mapping to Bloch spins.
Defining
$r_p^\pm=\la \frac12(\sigma_p^x\pm i\sigma_p^y)\ra$, $r_p^z=\la\sigma_p^z\ra$,
and rescaling $gb \to b$, we
write
\bea\label{eq:bloch1}
&& i\dot r_p^- = (p^2/m) r_p^- - b r_p^z
\,,\quad 
i\dot r_p^z = 2b r_p^+ - 2b^\ast r_p^-
\\\label{eq:bloch3}
&& i\dot b = \omega b + g^2\sum_p r_p^-
\eea
Since the norm is preserved by Bloch time evolution,
$|\vec r_p|^2 = 4 r_p^-r_p^+ + (r_p^z)^2$ is conserved for each 
spin.
We apply rotation,
\be
r_p^- \to e^{- i\eta t}r_p^-
\,,\quad
r_p^+ \to e^{i\eta t}r_p^+
\,,\quad
b \to e^{- i\eta t} b
\ee
with the value $\eta$ to be determined later.
This is equivalent to shifting
$p^2/m \to \epsilon_p=p^2/m - \eta$ and 
$\omega \to \omega - \eta$.

The resulting problem possesses real-valued solutions
which can be obtained from
the standard ansatz~\cite{McCall67}
\be\label{eq:ansatz}
r_p^-=A_pb+iB_p\dot b
\,,\quad
r_p^z=D_p-C_pb^2
\ee
Substituting this into 
Eqs.(\ref{eq:bloch1}),(\ref{eq:bloch3}), 
from the real part of the equation or $r_p^-$ and from the 
the equation or $r_p^z$ one 
finds the relations between
the ansatz parameters
$A_p=\epsilon_p B_p$,
$C_p=2 B_p$
while the imaginary part of the equation or $r_p^-$ generates
a set of equations
\be\label{eq:''b}
B_p\ddot b + \epsilon_p b - b(D_p-C_pb^2)=0
\ee
The constant of motion
$|\vec r_p|^2=4 r_p^-r_p^++(r_p^z)^2$ 
provides a first integral
of Eq.(\ref{eq:''b}):
\be\label{eq:b'^2_p}
4\lp \epsilon_p^2 b^2 + \dot b^2\rp + \lp 2b^2-D_p/B_p\rp^2 
= B_p^{-2}|\vec r_p|^2
\ee
where we expressed $A_p$ and $C_p$ through $B_p$.

\begin{figure}[t]
\centerline{
\begin{minipage}[t]{3in}
\vspace{0pt}
\centering
\includegraphics[width=3in]{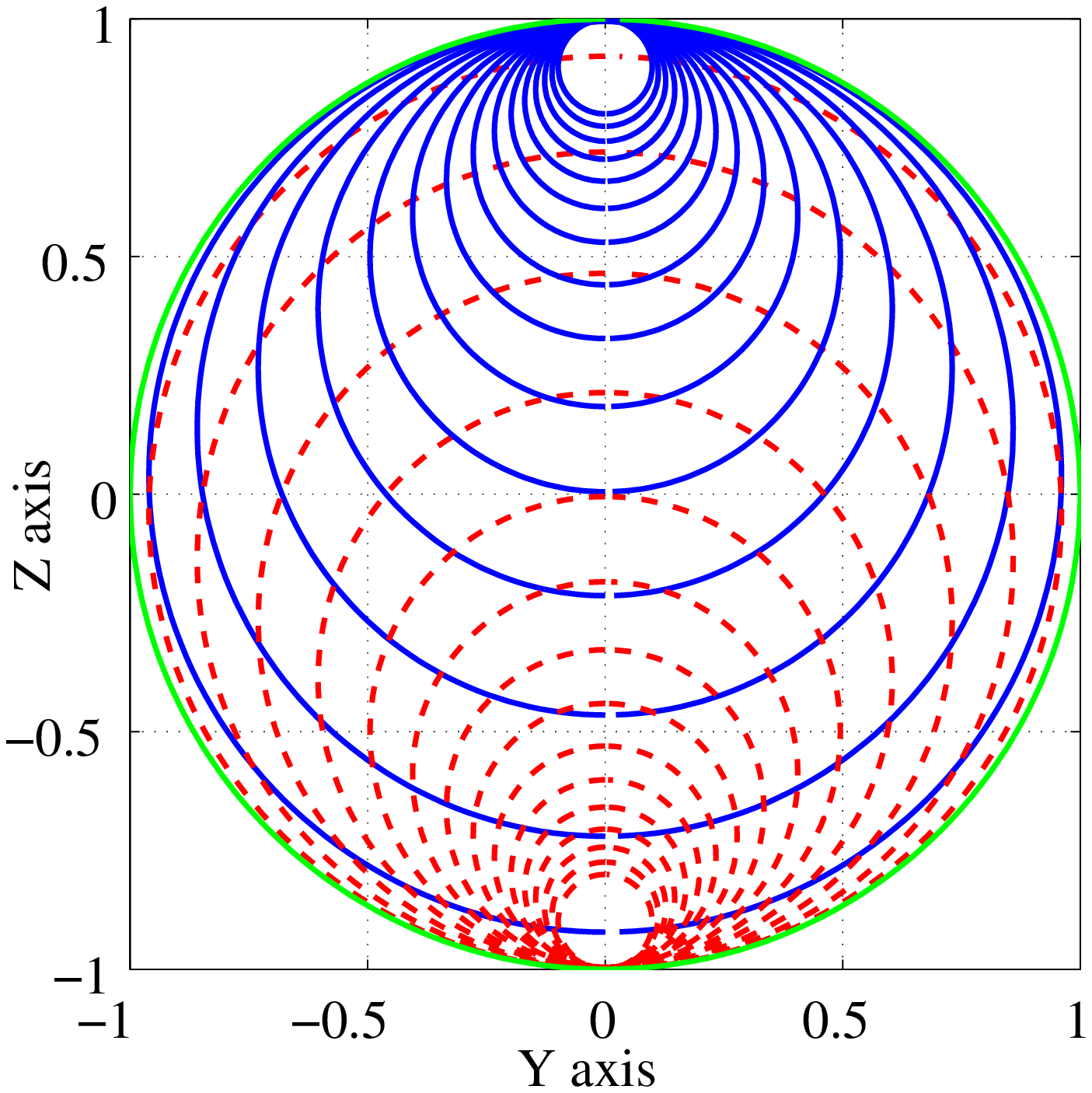}
\end{minipage}
\hspace{-1.9in}
\begin{minipage}[t]{1.9in}
\vspace{0.7in}
\centering 
\includegraphics[width=1.8in]{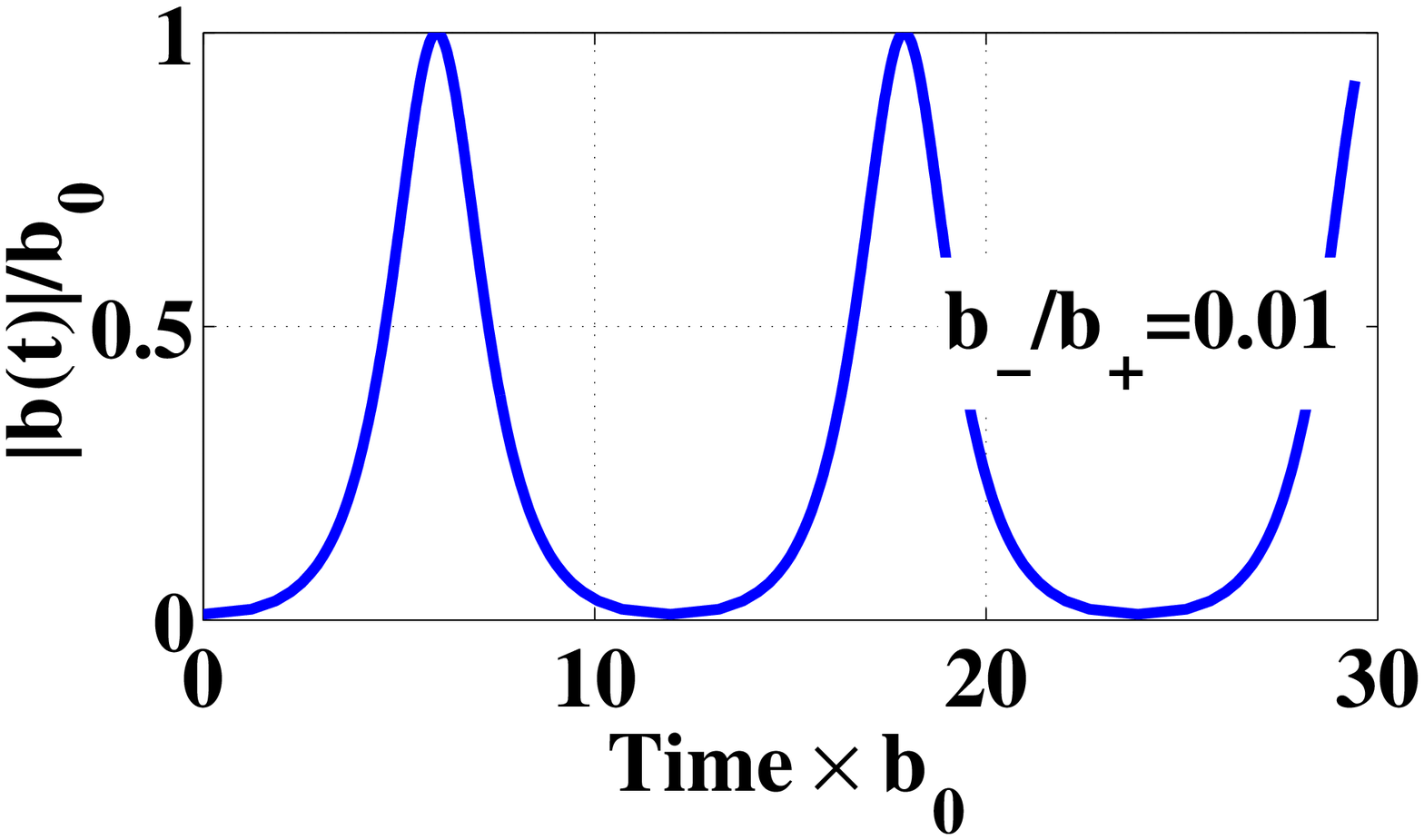}
\end{minipage}
} \vspace{0.15cm} \caption[]{
Time evolution (\ref{eq:ansatz})
of fermion pair amplitudes with different energies 
obtained for the molecular
field of a soliton train form (\ref{eq:cosh_soliton}),
shown in the inset.
The Bloch sphere parameterization
of pseudospin variables 
$r_p^{x,y,x}=\la \sigma_p^{x,y,x}\ra$ is used. 
Pseudospins precess so that
each state
completes a full $2\pi$ Rabi cycle per soliton. 
The red and blue curves
correspond to the energies
above and below the Fermi level. 
}
\label{fig:Bloch_sphere}
\end{figure}

Evidently, since the function $b(t)$ is the same for all spins,
the dependence on $p$ has to drop out of Eq.(\ref{eq:b'^2_p}),
giving a single equation for $b$ of the form
\be\label{eq:b'^2}
\dot b^2 = (b^2-b_-^2)(b_+^2-b^2)
\,,\quad b_-<b_+
\ee
which is possible with the following choice of constants:
\be
D_p/B_p - \epsilon_p^2 = b_-^2+b_+^2
\,,\quad
D_p^2-|\vec r_p|^2 = 4b_-^2b_+^2B_p^2
\ee
These equations determine the modulus of $B_p$ and $D_p$ only.
The sign has to be determined from initial conditions:
${\rm sgn}\,B_p={\rm sgn}\,D_p={\rm sgn}\,r_p^z$,

The solution of Eq.\,(\ref{eq:b'^2}) is
an elliptic function $b(t)={\rm dn}(t,\kappa^2)$ 
with $\kappa^2=1-b_-^2/b_+^2$~\cite{Ryzhik},
oscillating periodically between
$b_-$ and $b_+$. At $b_-\ll b_+$,
the solution is approximately given by a train
of weakly overlapping solitons 
\be\label{eq:cosh_soliton}
b(t)=\sum_n\frac{\gamma}{\cosh \gamma(t-t_n)}
\,,\quad t_n=\tau n 
\ee
(Fig.\,\ref{fig:Bloch_sphere}),
where each soliton in Eq.(\ref{eq:cosh_soliton}) is a solution
of Eq.\,(\ref{eq:b'^2}) with $b_-=0$, $b_+ =\gamma$.

The quantities $b_\pm$ 
and the frequency $\eta$ must be determined from the 
equation for $b$. One verifies that Eq.(\ref{eq:bloch3})
is consistent with the ansatz~(\ref{eq:ansatz}) and obtains
\bea\label{eq:self_consistency}
1&=& g^2\sum_p \frac{r_p^z}{\sqrt{\lp\epsilon_p^2+b_-^2+b_+^2\rp^2-4b_-^2b_+^2}},\\
\omega&=&\eta-g^2\sum_p\lp\frac{\epsilon_p r_p^z}{\sqrt{\lp\epsilon_p^2+b_-^2+b_+^2\rp^2-4b_-^2b_+^2}}
+\frac{1}{p^2/m}\rp,\nonumber
\eea
Here $r_p^z=2n_p-1$ corresponds to the energy distribution $n_p$
of fermions which depends on the initial state.
For the parameters used in Fig.~\ref{fig:Phase-diagram},
by order of magnitude we estimate
$\gamma,\tau^{-1} \simeq g^2 n/E_F \approx 8\,{\rm MHz}$.
This is much faster than typical energy relaxation rates,
which justifies ignoring relaxation effects in the dynamics.

The properties at equilibrium can be understood by
considering the limit $b_- \to b_+=b_0$,
when oscillations are absent. 
The energy distribution $n_p$ can be easily obtained 
in the pseudospin picture, taking into account that each spin
is presented with a tilted field 
$\vec h_p=(b_0,0,p^2/2m-\mu)$
which gives $n_p=1/\lp 1+ e^{\beta |\vec h_p|}\rp$.
The molecular field $b_0$ in the ground state
is determined by
\be\label{eq:equilibrium}
\omega = \eta+g^2\sum_p\lp\frac{{\rm sgn}\,\epsilon_p \lp1-2 n_p\rp}{\sqrt{\epsilon_p^2+4b_0^2}}
-\frac{1}{p^2/m}\rp
\ee
along with the 
constraint
$N=2b_0^2/g^2+2\sum_p n_p$.

Here we use Eq.(\ref{eq:equilibrium}) to verify the 
above stability analysis. 
To determine when the atoms can be stable with respect 
to hybridizing with
molecules, we set $b_0=0$ and immediately recover
Eq.(\ref{eq:lambda}) for the instability exponent 
$\lambda$. The difference, however, is that $\eta$ defined by Eq.(\ref{eq:equilibrium}) is real, while
$\lambda$ is complex. Atoms' stability is thus indeed equivalent to
the existence of a real-valued solution
of Eq.(\ref{eq:lambda}). 
One possibility to have such a solution
is to set $\eta=2\mu$, which eliminates the log divergence 
in (\ref{eq:equilibrium}) at $\epsilon_p=0$.
The other possibility is to have $\mu,\eta\le 0$. 
Put together with the properties
of equilibrium state at finite $b_0$, 
this confirms the above 
estimate of the coexistence region (\ref{eq:boundary})
and the conclusion that pure atom state is metastable at 
the detuning
$\omega<\omega_0$. 

In summary, this work provides a phase digram
and an exact solution for
the atom-molecule dynamics in the regime of strong coupling.
The characteristic energy scales are estimated 
to be much larger than 
$E_F$, which makes the Dicke model approximation ignoring
molecular dispersion as well as the BCS fermion pairing
effects, accurate enough. A wide atom-molecule
coexistence region is predicted in which atom pairs 
and molecules
hybridize into objects of size much less
than Fermi wavelength $p_F^{-1}$.

After having completed this work 
we became aware of the article by 
Andreev, Gurarie and Radzihovsky\cite{andreev04}
which exploits the mapping to the Dicke model, while
focusing on the weak coupling limit.

\end{multicols}
\end{document}